\documentclass[preprint,superscriptaddress,twocolumn,lengthcheck,tightenlines,showpacs,a4paper,10pt,prl]{revtex4}
\usepackage{amssymb}
\usepackage{amsfonts}
\usepackage{color}
\usepackage{graphicx}

\input{tcilatex}
\begin{document}

\preprint{APS/123-QED}
\title{Contiguous $3d-$ and $4f-$magnetism:\\
towards strongly correlated $3d-$electrons in YbFe$_2$Al$_{10}$}
\author{P. Khuntia*}
\affiliation{Max Planck Institute for Chemical Physics of Solids, 01187 Dresden, Germany}
\author{P. Peratheepan}
\affiliation{Highly Correlated Matter Research Group, Physics Department, University of
Johannesburg, P.O. Box 524, Auckland Park 2006, South Africa}
\affiliation{Department of Physics, Eastern University, Vantharumoolai, Chenkalady 30350,
Sri Lanka}
\author{A. Strydom}
\affiliation{Max Planck Institute for Chemical Physics of Solids, 01187 Dresden, Germany}
\affiliation{Highly Correlated Matter Research Group, Physics Department, University of
Johannesburg, P.O. Box 524, Auckland Park 2006, South Africa}
\author{Y. Utsumi}
\affiliation{Max Planck Institute for Chemical Physics of Solids, 01187 Dresden, Germany}
\author{K.-T. Ko}
\affiliation{Max Planck POSTECH Center for Complex Phase Materials, 01187 Dresden,
Germany and Pohang 790-784, Korea}
\author{K.-D. Tsuei}
\affiliation{National Synchrotron Radiation Research Center, 101 Hsin-Ann Road, Hsinchu
30077, Taiwan}
\author{L. H. Tjeng}
\affiliation{Max Planck Institute for Chemical Physics of Solids, 01187 Dresden, Germany}
\author{F. Steglich}
\affiliation{Max Planck Institute for Chemical Physics of Solids, 01187 Dresden, Germany}
\author{M. Baenitz}
\affiliation{Max Planck Institute for Chemical Physics of Solids, 01187 Dresden, Germany}

\begin{abstract}
We present magnetization, specific heat, and $^{27}$Al NMR investigations on
YbFe$_{2}$Al$_{10}$ over a wide range in temperature and magnetic field. The
magnetic susceptibility at low temperatures is strongly enhanced at weak
magnetic fields, accompanied by a $\ \ln (T_{0}/T)$ divergence of the low$-T$
specific heat coefficient in zero field, which indicates a ground state of
correlated electrons. From our hard X-ray photoemission spectroscopy
(HAXPES) study, the Yb valence at 50 K is evaluated to be 2.38. The system
displays valence fluctuating behavior in the low to intermediate temperature
range, whereas above 400 K, Yb$^{3+}$ carries a full and stable moment, and
Fe carries a moment of \ about 3.1 $\mu _{B}.$ The enhanced value of the
Sommerfeld Wilson ratio and the dynamic scaling of spin-lattice relaxation
rate divided by \textit{T} $\ [^{27}$($1/T_{1}T)]$ with static
susceptibility suggests admixed ferromagnetic correlations. $^{27}$($%
1/T_{1}T)$ simultaneously tracks the valence fluctuations from the 4\textit{f%
} -Yb ions in the high temperature range and field dependent
antiferromagnetic correlations among partially Kondo screened Fe 3\textit{d}
moments at low temperature, the latter evolve out of an Yb 4\textit{f }
admixed conduction band.
\end{abstract}

\volumeyear{year}
\volumenumber{number}
\issuenumber{number}
\eid{identifier}
\date[Date text]{date}
\received[Received text]{date}
\revised[Revised text]{date}
\accepted[Accepted text]{date}
\published[Published text]{date}
\startpage{1}
\endpage{2}
\maketitle


Novel phases ranging from unconventional superconductivity and spin liquid
to quantum criticality in correlated electron systems result from competing
interactions between magnetic, charge, orbital and lattice degrees of
freedom \cite{SS,FS}. Competing interactions such as the mostly
antiferromagnetic (AFM) Rudermann-Kittel-Kasuya-Yosida (RKKY) exchange and
the Kondo effect on a localized spin may lead to a magnetic instability
which generates unusual temperature ($T$) and magnetic field ($H$) scaling
behavior of bulk and microscopic observables. The competing magnetic
interactions frequently produce generalized non-Fermi liquid (nFL) scaling
in the thermal behavior of physical properties. If the RKKY spin exchange
succeeds in overcoming the thermal energy of the spin system conducive to a
paramagnetic-to-AFM transition, the addition of a competing Kondo spin
exchange with the conduction electrons achieves a curbing effect on the
phase transition. Moreover, under favorable conditions such as applied
pressure or magnetic field the phase transition may become confined to
temperatures arbitrarily close to zero, which in turn leads to remarkable
thermal scaling in the realm of quantum criticality\cite{RMP1,RMP2,LH,MB,QS}%
. In exceptional cases quantum criticality presents itself under ambient
conditions, such as in U$_{2}$Pt$_{2}$In \cite{ams96,estrela99} or in the
superconductor $\beta -$YbAlB$_{4}$ \cite{SN}. Quantum criticality stemming
from ferromagnetic exchange on the other hand is a rare occurrence, and has
been discussed among $5f-$electron systems such as UGe$_{2}$ \cite{SSC,HK}
or UCoGe\cite{ES,TH}, $4f$ systems like YbNi$_{4}($P$_{1-x}$As$_{x}$)$_{2}$ 
\cite{RS1,AS}, Ce(Ru$_{1-x}$Fe$_{x}$)PO\cite{SK1,SK2}, and in weak itinerant
ferromagnets like ZrZn$_{2}$ \cite{PRS} and NbFe$_{2}$ \cite{MB2}. YFe$_{2}$%
Al$_{10}$, an isostructural version of YbFe$_{2}$Al$_{10}$ with no $4f-$%
electrons, is reported to be a plausible candidate for a FM quantum critical
magnet\cite{PK,AM,per10,KP,str13}.

The ternary orthorhombic aluminides of $RM_{2}$Al$_{10}$ type ($R=$rare
earth element, $M=$Fe, Ru, Os) have been the subject of considerable debate
in view of a fascinating conundrum of physical properties. Most notable are
the extremes of magnetic interactions found in the Ce series ranging from
unprecedently high AFM order at $27~$K in CeRu$_{2}$Al$_{10}$ \cite%
{CS,SC,AM1,DD} to the Kondo insulating state in CeFe$_{2}$Al$_{10}$ \cite{YM}%
. In the present study to further unravel the nature of the $3d-$electrons
in this class of material, we assess the response of Fe-based magnetism in
the presence of localized magnetism, namely the rare earth element Yb, and
we use a combination of bulk and microscopic probes due to the anticipated
complexity of an admixture of different types of magnetic exchange. A
comparable situation can be found in CeFe$_{2}$Al$_{10}$ in which the
confluence of the two types of magnetic species has the surprising effect of
producing the non-magnetic Kondo insulating state \cite{YM}, which is an
extreme case of local-moment hybridization with the conduction electrons.
Recently, there has been a resurgence of research activities in intermediate
valence systems following the discovery of superconductivity and quantum
critical behavior in an intermediate valence (IV) heavy fermion $\beta -$%
YbAlB$_{4}$\cite{SW,DTA,SP,SN,PC,MO,PC1,LMH,WS}.

\ In this Letter, we present comprehensive magnetic susceptibility, specific
heat, and $^{27}$Al NMR investigations on polycrystalline YbFe$_{2}$Al$_{10}$%
.\ Furthermore, hard X-ray photoemission spectroscopy (HAXPES) at SPring-8,
Japan was carried out as a direct probe of the valence state of Yb. Magnetic
susceptibility and specific heat display low temperature divergences, yet
without any signature of magnetic ordering down to $0.35~$K. In order to
understand the low energy spin dynamics governing the underlying magnetism
of the title compound, we have carried out NMR investigations with special
attention to the spin-lattice relaxation measurements. The low field
spin-lattice relaxation rate shows a divergence towards low temperatures,
which is consistent with magnetization and specific heat data. The observed
deviations from the FL behavior is associated with correlated $3d$ Fe
moments strongly coupled via the conduction band, which is hybridized with
the Yb derived 4$\mathit{f}$ states.

Polycrystalline samples of YbFe$_{2}$Al$_{10}$ have been sythesized
following a method discussed elsewhere\cite{AM,AM1}. The \textit{dc}
magnetic susceptibility $\chi (T)$ (=$M(T)/H)$ and thermopower data were
obtained using a QD PPMS.\newline
In a recent work, a Kondo-like electrical resistivity accompanied by
divergences in magnetic susceptibility, $\chi (T)$ and the Sommerfeld
coefficient $\gamma $($T$) = C$_{p}$(\textit{T})/\textit{T }(where C$_{p}$($%
T $) is the electronic specific heat) in zero field were reported \cite%
{per10} on YbFe$_{2}$Al$_{10}$. The magnetism of Yb in this compound was
demonstrated \cite{PKSS} to be subject to an unstable valence and to recover
its full trivalent state at $T$ \TEXTsymbol{>} 400 K, which is in agreement
with earlier reports\cite{MV}.

Shown in Fig.\ 1 (a) is the field dependent magnetic susceptibility, $\chi
(T)$ of YbFe$_{2}$Al$_{10}$. The values of $\chi (T)$ are enhanced by one
order of magnitude in comparison with the non-$4f$ electron homologue YFe$%
_{2}$Al$_{10}$ \cite{PK}, which indicates a strong hybridization of the Yb
derived $4f$ states with the conduction electron states. A modified band
structure is therefore expected with subsequent effects on the itinerant $3d$
magnetism of Fe. Towards elevated temperatures, Yb tends to reach its full
trivalent state and this temperature-driven evolution is appropriately
reflected in the thermopower $S(T)$ \textit{i.e}., by a broad peak centered
at $T$* ($\approx $100 K) (see Fig.\ 1(d)), which we use to denote the
temperature scale of the valence change of Yb. Such a peak in thermopower is
typical for IV Yb compounds, and our HAXPES study on YbFe$_{2}$Al$_{10}$
confirmed such an IV state of Yb in this compound and valence of Yb is
evaluated to be 2.38 (see Supplemental Material for more details). At the
high temperature end, the consequence of this peak is played out by a change
in the sign of $S(T)$ at about $300~$K, which likely implies a
temperature-driven change in the relative weights and participation of both
holes and electrons in the underlying bandstructure. This could also be due
to the asymmetry of the density of states (DOS) or the scattering rate at
the Fermi energy for a single band. However, the negative sign in $S(T)$ of
YbFe$_{2}$Al$_{10}$ signals stable and local-moment magnetic character of Yb
above 300 K, because \textit{S}($T$) native to the weakly hybridized $%
4f^{13+\delta }$ state of Yb is expected to be negative \cite{beh04}.

The small upturn in S($T$) below 10 K is consistent with the incoherent
Kondo like restistivity $\rho (T)$\cite{per10}. A peak in $\rho (T)$ at 
\textit{T} $\simeq $ 4.5 K (see Supplemental) is reminiscent of
Kondo-lattice behavior\cite{per10}.The Kondo type upturn in $\rho (T)$ as
well as the low$-T$ divergence in $\chi (T)$ is quenched by applying
magnetic fields of a few teslas. The initial susceptibility $\chi
(H\rightarrow 0)$ at $2~$K as well as the high-field magnetization $M(H)$
yield extremely small values of the magnetic moment in YbFe$_{2}$Al$_{10}$.
Following a weak curvature in the M($H$) in low fields, there is however no
saturation achieved in $M(H)$ at $2~K$ even up to 7 T , Fig.\ 1(c), where a
quasi-linear in field magnetization is found.\textbf{\ }


\FRAME{fhFU}{3.3088in}{2.655in}{0pt}{\Qcb{(Color online) Temperature
dependence of $\protect\chi $ in various applied magnetic fields (b) $1/%
\protect\chi $ \emph{vs}.\ $T$ at $0.1~$T with Curie-Weiss fits. (c)
Magnetization isotherm at $2~$K and the inset shows the $\protect\chi ^{-1}$
vs. \textit{T} in 5 kOe and 10 kOe with Curie Weiss fit as discussed in the
text (d) Temperature dependence of thermopower.}}{}{fig1.eps}{\special%
{language "Scientific Word";type "GRAPHIC";maintain-aspect-ratio
TRUE;display "ICON";valid_file "F";width 3.3088in;height 2.655in;depth
0pt;original-width 11.3757in;original-height 7.9468in;cropleft "0";croptop
"1";cropright "0.8719";cropbottom "0";filename '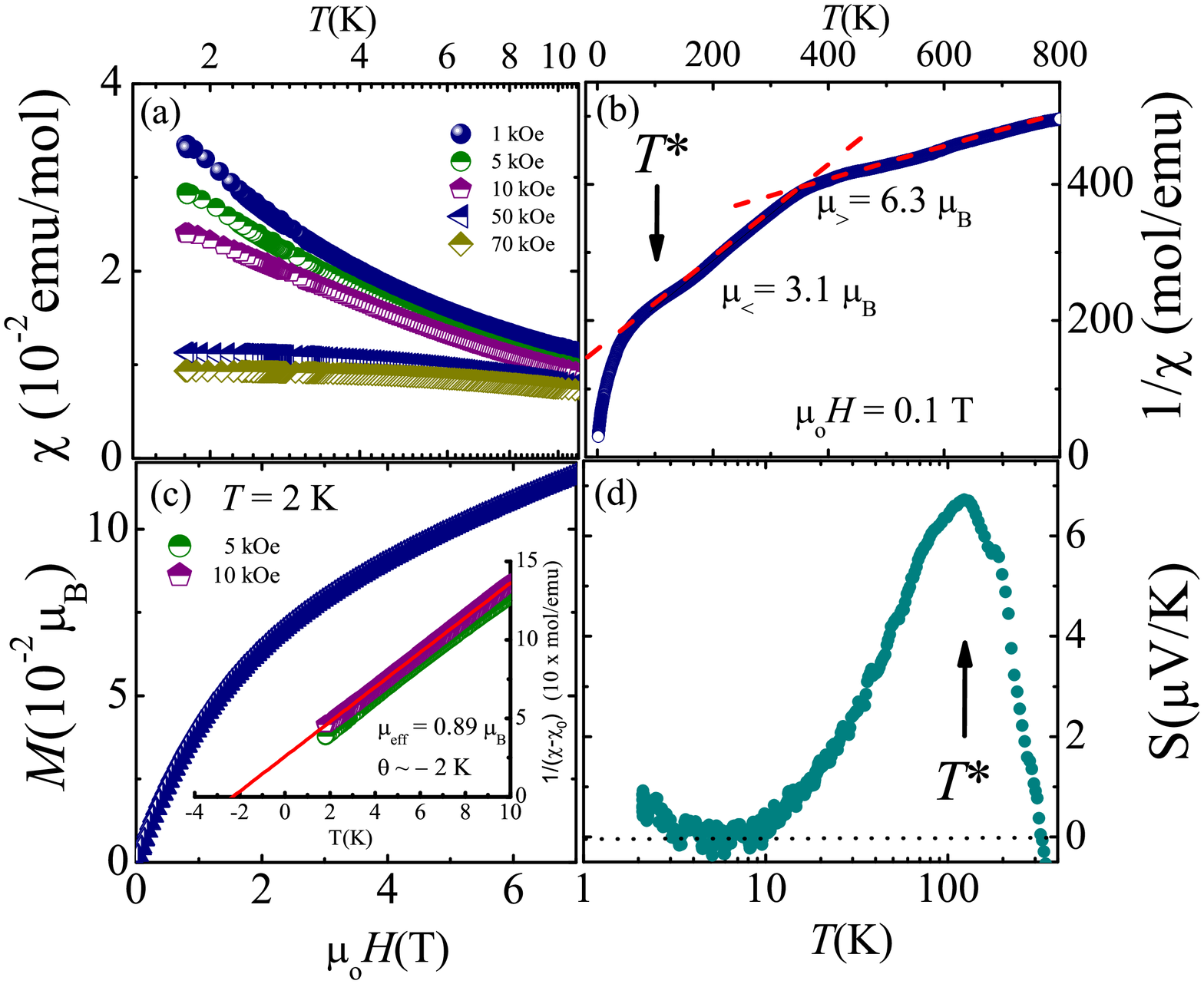';file-properties
"XNPEU";}}


A detailed analysis of $\chi (T)$ (see Fig 1b) reveals an intermediate
valence (IV) state of Yb at low and intermediate temperatures, but Yb
recovers its full moment (4.54 $\mu _{B}$) with a high spin state Fe (3.1 $%
\mu _{B}$) at $T>$400 K with predominant AFM correlations. A similar
scenario has been discussed in the IV Yb-based skutterudites YbFe$_{4}$Sb$%
_{12}$ \cite{sch05} and YbFe$_{4}$P$_{12}$\cite{AY}. The deconvolution of
the Fe$-3d$ contribution and the more localized Yb$-4f$ contribution is a
daunting task and beyond the scope of this manuscript. Nonetheless, based on
the model of Rajan\cite{VTR}, for the Yb$-4f$ part a constant and $T$%
-independent susceptibility could be expected towards low temperatures. The
HAXPES measurement performed with \textit{h}$\upsilon $=6.5 keV at 50 K
confirms the IV state of Yb with valence 2.38 (see Supplemental). Therefore,
we assume that the magnetism below 50 K is solely driven by the Fe \ 3$d$
moments in YbFe$_{2}$Al$_{10}$ and we speculate that the Curie-Weiss
behavior of $\chi (T)$ in the intermediate temperature range 80$\leq $%
\textit{T}$\leq $370 K is associated to Fe-3\textit{d} moments with an
effective moment of 3.1/$\sqrt{2}$= 2.2 $\mu _{B}.$This is in contrast to
its non-4$f$ analog YFe$_{2}$Al$_{10}$ where Fe carries a much smaller
magnetic moment of 0.35 $\mu _{B}$ per Fe\cite{PK,KP}. This might be related
to the difference in charge transfer from the divalent Yb$^{2+}$ to the Fe$%
_{2}$Al$_{10}$ host lattice in comparision to the trivalent Y$^{3+}$. The
Curie-Weiss behavior of $\chi (T)$\ at $T\leq $10 K (see Fig. 1(c) inset)
reveals a small Fe moment 0.89/$\sqrt{2}$=0.63 $\mu _{B}$ per Fe in YbFe$%
_{2} $Al$_{10}.$ The Weiss temperature of $\theta \simeq $ --2 K yields an
on-site Kondo temperature of the Fe moments, which amounts to several Kelvins%
\cite{AH}.

Shown in Fig. 2(a) is the specific heat coefficient $\gamma $($T$) in
different magnetic fields measured using the $^{3}$He option of QD PPMS. The
specific heat coefficient ($\gamma )$ is enhanced towards low temperatures
and follows a $\ln (T_{0}/T)$ behavior with $T_{0}$= 2 K in zero field,
which suggests a correlated behavior of electrons. This may be attributed to
entropy of unquenched spin degrees of freedom, or to impending cooperative
behavior at much lower temperatures. Applied magnetic fields achieve a
suppression and eventual saturation into a constant value of $C_{\mathrm{p}%
}/T$ and thus the recovery of the Fermi liquid ground state. The ratio of
the enhanced $\gamma _{0}$ value at zero field to the fully quenched value $%
\gamma _{\mathrm{H}}$ in $9~$T at $0.35~$K is about $2.5$. Surprisingly,
this enhancement factor is qualitatively similar to that of the non-$4f$
compound YFe$_{2}$Al$_{10}$ \cite{PKSS}. Despite the fact that the relative
enhancements $\Delta \gamma /\gamma _{H}=(\gamma _{0}-\gamma _{H})/\gamma
_{H}$\ are similar, it should be mentioned that the $T$ dependencies of \ C$%
_{p}$/$T$ \ are dissimilar: (ln($T_{\mathit{0}}$\textit{/}$T$\textit{)} for
YbFe$_{2}$Al$_{10}$ and power law behavior in case of YFe$_{2}$Al$_{10}$).
For YbFe$_{2}$Al$_{10}$ the magnetic entropy (0.014\textit{R}ln2) below 2 K
is about three times larger than its non-4\textit{f} counter part YFe$_{2}$Al%
$_{10}$\cite{ESR}. Therefore, we relate the low-temperature divergence of
the Sommerfeld coefficient to the emergence of correlations among Fe moments
amplified by the strong hybridization between Yb$-4f$ states and $s+d$
conduction band states at the Fermi level. The field dependence of the
Sommerfeld coefficient at $0.5~$K follows a $H^{-0.35}$ behavior (Fig.\ 2b),
and the transition to a constant in temperature regime provides the
crossover scale between FL and nFL behavior (inset of Fig.\ 2b).


\FRAME{fhFU}{3.6357in}{2.1819in}{0pt}{\Qcb{(Color online) (a) Temperature
dependence of specific heat coefficients taken in different applied magnetic
fields. The solid line is a fit to $\ln (T_{0}/T)$ with \textit{T}$_{0}$ = 2
K. (b) $C_{\mathrm{P}}(T)/T$ \emph{vs.}\ $\protect\mu _{0}H$ at $0.5~$ and $%
2~$K, with solid line is a fit to \textit{H}$^{-0.35}$. The inset shows the
field dependence of the cross over temperature (to FL behavior) obtained
from $C_{\mathrm{P}}(T)/T$.}}{}{fig2.eps}{\special{language "Scientific
Word";type "GRAPHIC";maintain-aspect-ratio TRUE;display "ICON";valid_file
"F";width 3.6357in;height 2.1819in;depth 0pt;original-width
4.5939in;original-height 2.9499in;cropleft "0.0433";croptop
"0.9082";cropright "0.9539";cropbottom "0.0608";filename
'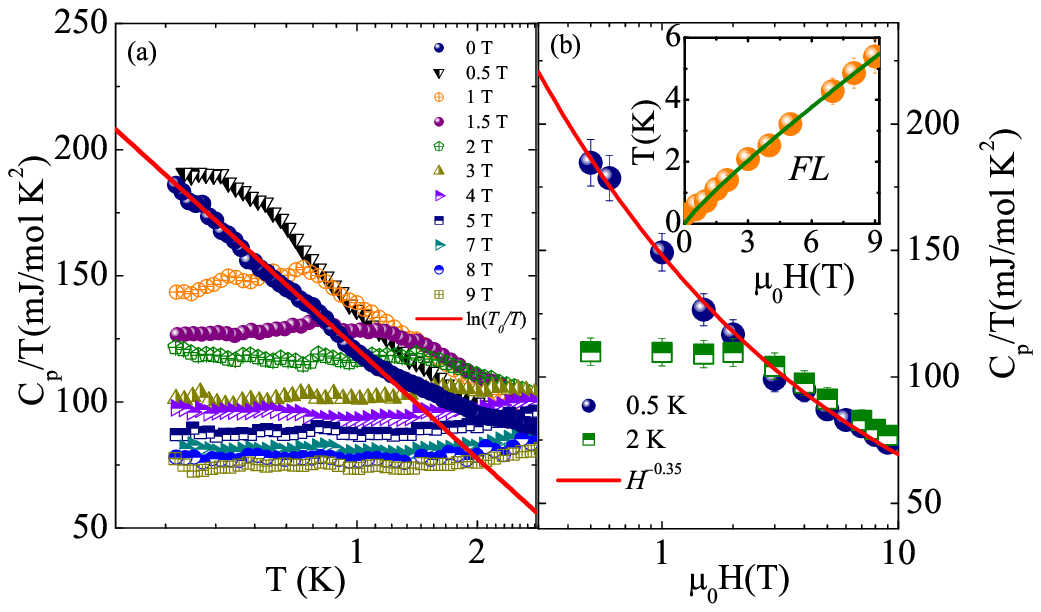';file-properties "XNPEU";}}


The residual quenched Sommerfeld coefficient of $\gamma _{H}$ = 75.3~mJ/mol K%
$^{2}$ in YbFe$_{2}$Al$_{10}$ exceeds that of the La equivalent \cite{YM} by
a factor $\sim 3$, which indicates that the Fermi level in YbFe$_{2}$Al$%
_{10} $ is occupied predominantly by heavy charge carriers. An enhanced
value of the Sommerfeld-Wilson ratio $R_{\mathrm{W}}=\pi ^{2}k_{\mathrm{B}%
}^{2}/\mu _{0}\mu _{\mathrm{eff}}(\chi /\gamma )\approx 12$ at $2~$K
indicates the presence of FM correlations. It is worth to mention that there
is a striking similarity of YbFe$_{2}$Al$_{10}$ specific heat data shown in
Fig.\ 2a to those of $\beta -$YbAlB$_{4}$ \cite{SN}, which is a rare example
of an IV system with local moment low-\textit{T} electron correlations\cite%
{LMH}. Another prominent example in that context is the IV metal YbAl$_{3}$%
\cite{ZF}.

$^{27}$Al-NMR ($I=5/2$) measurements have been performed using a standard 
\emph{Tecmag} NMR spectrometer in the temperature range $1.8\leq T\leq 300~$%
K and in the field range $0.98\leq \mu _{0}H\leq 7.27~$T. The orthorhombic
crystal structure of YbFe$_{2}$Al$_{10}$ hosts five inequivalent Al sites.
Usually this results in rather broad NMR spectra with a clear central
transition and superimposed first order satellite transitions. Surprisingly,
we found a rather well-resolved central transition with a small field
dependent anisotropy, which implies that the different Al sites are rather
equal in their magnetic environment\cite{PK,PKSS}.


\FRAME{fhFU}{3.2171in}{2.4552in}{0pt}{\Qcb{(Color online) $^{27}$Al NMR
spectra at $80~$MHz for different temperatures.The inset shows the
simulation of the $4.3~$K spectra.}}{}{fig3.eps}{\special{language
"Scientific Word";type "GRAPHIC";maintain-aspect-ratio TRUE;display
"ICON";valid_file "F";width 3.2171in;height 2.4552in;depth
0pt;original-width 4.0101in;original-height 3.3382in;cropleft
"0.0447";croptop "0.8802";cropright "0.9327";cropbottom "0.0686";filename
'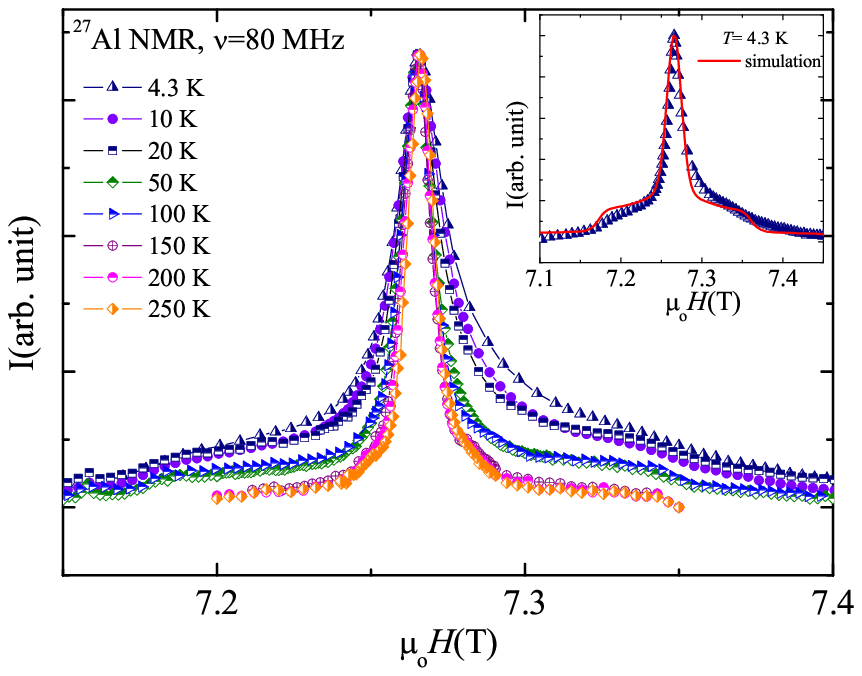';file-properties "XNPEU";}}


The sharp central transition enables us to perform $^{27}$Al spin-lattice
relaxation rate (SLRR) measurements consistently following saturation
recovery method with suitable \emph{rf} pulses and the results are shown in
Fig.4. The relaxation rate divided by $T$, \emph{i.e.} $1/T_{1}~T$, shows a
divergence towards low temperatures (Fig. 4a) with a proportionality $\chi
(T)/\sqrt{T}$ in the lowest magnetic fields. Such a dynamic scaling is
frequently found in heavy fermion systems with AFM correlations and even
with admixed FM correlations like in CeFePO \cite{DC,NB,MB1}. In addition,
the relative change $\left[ (\gamma _{0}-\gamma _{\mathrm{H}})/\gamma _{%
\mathrm{H}}\right] ^{2}=1.7$ underestimates the SLRR enhancement found in
the experiment ($\simeq 4.6$). The stronger enhancement in the SLRR points
towards the presence of dominant $q=0$ contributions, as a response to FM
correlations. Usually the specific heat is more sensitive to finite $q$
excitations which explains the difference in the enhancement factors. In
contrast with the discrepancy in the $T$ enhancement, the field dependence
of the SLRR is in agreement with the Fermi liquid theory exhibiting $%
1/T_{1}T\sim (C/T)^{2}\sim \gamma ^{2}$ behavior. Here a power law $%
H^{-0.35} $ is found for the Sommerfeld coefficient which implies a power
law $H^{-0.7} $ for the SLRR. This field dependence is indeed found for the
SLRR at $2.5~$K (see Fig.\ 4b with \ $\mathit{H}^{-0.77}$), which is further
supported by $1/T_{1}T\sim \chi ^{2}$ behavior commonly found in
local-moment metals and is in contrast to that observed in YFe$_{2}$Al$_{10}$
\cite{PK}.\newline
\ Independent of the magnetic field a peak (Fig.4a) in the SLRR at $T^{\ast
} $($\simeq $100 K) signals the onset of valence fluctuations in the SLRR of
the Al nuclei at high temperature. In general the SLRR probes the $q-$%
averaged low lying excitations in the spin fluctuation spectra and $1/T_{1}T$
can be expressed as; $\frac{1}{T_{1}T}\propto \sum_{q}\mid A_{hf}(q)\mid ^{2}%
\frac{\chi ^{\prime \prime }(q,\omega _{n})}{\omega _{n}}$ ,where $A_{hf}(q)$
is the $q-$dependent form factor of the hyperfine interactions and $\chi
^{\prime \prime }(q,\omega _{n})$ is the imaginary part of dynamic electron
spin susceptibility \cite{TM,YK}. In the presence of $q-$isotropic $4f$
fluctuations of IV Yb coexisting with $q=0$ FM $3d$ correlations, the SLRR
could be approximated by $\frac{1}{T_{1}T}\simeq A_{hf}{}^{2}\chi _{0}\left[
\tau _{3d}+\tau _{4f}\right] $, where $\tau _{4f}(=1/\Gamma _{\mathrm{4f}}$)
is the effective fluctuation time of the $4f$ ion, $\tau _{3d}(=1/\Gamma _{%
\mathrm{3}\mathit{d}}$ ) is the effective fluctuation time of the $3d$ ion ($%
\Gamma _{4f},_{3d}$ \ are corresponding dynamic relaxation rates), and $\chi
_{0}$ is the uniform bulk suceptibility. It has to be mentioned that in case
of large valence variations (like in Eu systems where the valence could vary
between 2+ and 3+) the electronic structure may be perturbed which changes $%
A_{\mathrm{hf}}$, but we omit this detail for YbFe$_{2}$Al$_{10}$ and assume
that $A_{\mathrm{hf}}$ is not varying with temperature.


\FRAME{fhFU}{3.659in}{2.7389in}{0pt}{\Qcb{(Color online) (a)$^{27}$($%
1/T_{1}T $ ) \emph{vs.}\ $T$ in different applied magnetic fields. The solid
line is the calculated value as discussed in the text. (b) The field
dependence of $^{27}$($1/T_{1}T)$ at 2.5 K with a fit to \textit{H}$^{-0.77}$%
.(c) The temperature dependence of \ $\protect\tau _{4f}$ at 7.27 T. }}{}{%
fig4.eps}{\special{language "Scientific Word";type
"GRAPHIC";maintain-aspect-ratio TRUE;display "ICON";valid_file "F";width
3.659in;height 2.7389in;depth 0pt;original-width 11.3757in;original-height
7.9468in;cropleft "0.0088";croptop "0.9064";cropright "0.6920";cropbottom
"0.1768";filename '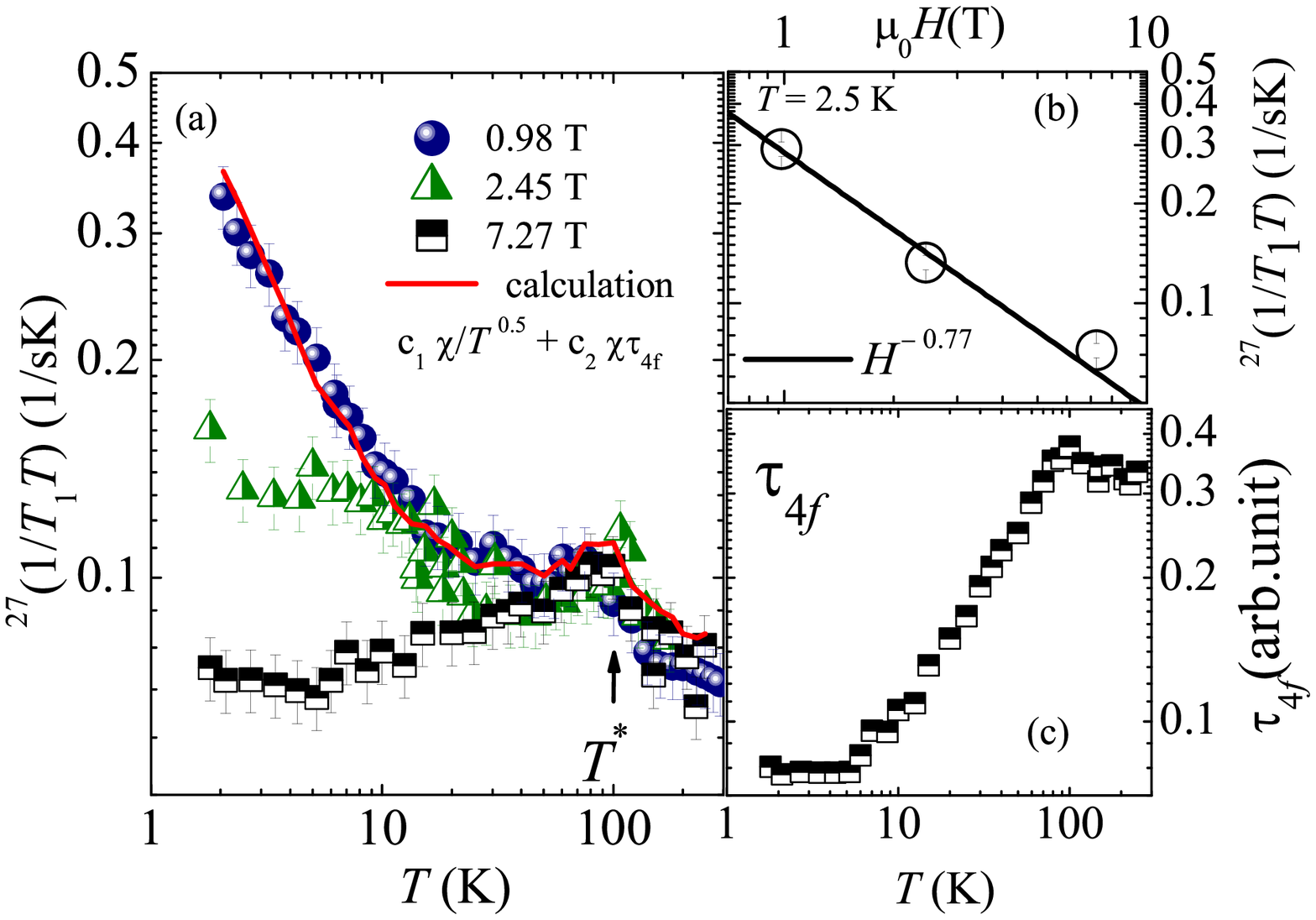';file-properties "XNPEU";}}


The beauty of these results is that $^{27}$ Al NMR simultaneously senses the
valence fluctuations from the 4$\mathit{f}$- Yb ions in the high temperature
range and the low temperature field dependent Kondo-like correlations
associated to the 3$\mathit{d}$- Fe ions. Upon the application of high
magnetic fields these fluctuations are quenched (here $\tau _{3d}\ll $ $\tau
_{4f}$ for entire temperature range). Therefore, the relaxation rate at 7.27
T allows for the determination of the effective fluctuation time $\tau
_{4f}=1/\Gamma _{4f},$which is plotted as a function of temperature in Fig.
4(c). The step like change of $\tau _{4f}$\ at about \ 100 K signals more a
charge gap scenario (like in Kondo insulators) than an intermediate valence
system with a smooth variation in $\tau _{4f}$. With the knowledge of the 
\textit{T} dependent (but not $\mathit{H}$-dependent) relaxation time $\tau
_{4f},$\ we now proceed to fit the 1/$T_{1}T$ vs. $T$ results in low
magnetic field. Surprisingly, the assumption\ of $\tau _{3d}=1/\sqrt{T}%
\propto \chi $ \ results in a very good agreement with the experimental data
(red line in Fig. 4(a)), which also explains the $^{27}$(1/$T_{1}T$\textit{)}%
$\propto \chi ^{2}$\ behavior at $T\rightarrow $0 limit. With this approach
we have convincingly shown that NMR is able to probe both energy regimes; i)
the high-temperature IV regime where $\Gamma _{\mathrm{4f}}$ is changing
strongly and ii) the low-\textit{T} regime where $\Gamma _{\mathrm{4f}}$ is
constant and $\Gamma _{\mathrm{3d}}$ shows a local moment behavior with $%
\Gamma _{\mathrm{3d}}\propto \sqrt{T}$.

In conclusion, we have found an unexpected localization of Fe-derived 3$d$
states upon cooling YbFe$_{2}$Al$_{10}$ to helium temperatures. As in this
material the Yb-derived 4$f$ electrons form a non-magnetic,
intermediate-valent state at low temperatures (with Yb valence 2.38), the
observed Kondo-lattice behavior has to be attributed to the localized 3$d$
electrons. Because of the low on-site Kondo scale of $\mathit{T}_{0}$ $%
\approx $ 2 K, one would expect the 3$d$ magnetic moments to be subject to
some kind of long-range ordering\cite{MI}. However, this appears to be
avoided, at least above 0.4 K, by a competition between ferro- and
antiferromagnetic correlations, which have been inferred from a strongly
enhanced Sommerfeld Wilson ratio on the one hand and the field dependencies
of the specific heat and spin-lattice relaxation rate on the other.

We thank C. Geibel, A. P. Mackenzie, H. Yasuoka, M. C. Aronson, M. Brando,
and M. Garst for fruitful discussions. We thank C. Klausnitzer for technical
support concerning specific heat measurements. We thank the DFG for
financial support (project OE-511/1-1). AMS acknowledges support from SA-NRF
(78832).

*pkhuntia@gmail.com

\subsection{I. \textbf{Magnetic susceptibility}}

The temperature dependence of the \textit{dc} magnetic susceptibility $\chi $%
($T$) was measured in different magnetic fields in the temperature range 1.8$%
\leq T\leq 300$ K using Quantum Design (QD) SQUID magnetometer. In addition,
measurements up to 800 K were carried out using the SQUID-Vibrating Sample
Magnetometer.

The Curie-Weiss (CW) fit of the inverse magnetic susceptibility 1/$\chi $($T$%
) in the temperature range 80 $\leq T\leq $ 370 K yields an effective Fe
moment of $\mu _{3d}$ =3.1/$\sqrt{2=}$ 2.2 $\mu _{B}$ whereas the CW fit in
the high temperature range 400 $\leq $ $T\leq $ 800 K results in an
effective moment of 6.3 $\mu _{B}$ which is too large to assign to Yb$^{3+}$
exclusively (Fig. 1b of the manuscript). This is due to the combined role of
Fe and Yb magnetic moments on the underlying magnetism of this system. Hence
we associate this net magnetic moment to arise from two Fe (two identical
octahedral sites occupied by Fe in the lattice) and one Yb moments. This
scenario is most likely in view of the intermediate valence state of Yb in
YbFe$_{2}$Al$_{10}$, which is confirmed by more definitive probe for valence
transition \textit{i.e}., HAXPES and the results are discussed in the
following section. The negative values of the Weiss temperatures obtained
from the CW fit in both temperature ranges suggest the dominant
antiferromagnetic correlations\cite{PK1,VMT}. The observed magnetic
susceptibility is strikingly different from the recently reported non-4$%
\mathit{f}$ homologue critical ferromagnet YFe$_{2}$Al$_{10}$ \cite{PK2}. In
addition the CW fit of the magnetic susceptibility data in the \ $T\leq $
10K unveil a very low Weiss temp, $\theta _{CW}\simeq $ -- 2 K, which
signals an on-site Kondo screening between Fe 3$\mathit{d}$ moment with
Kondo temperature of a few Kelvin. Furthermore, an evaluation of the
stability of the FM phenomena in YFe$_{2}$Al$_{10}$ revealed that the
quantum criticality found in YFe$_{2}$Al$_{10}$ is not pliant with small
variation in the Fe content\cite{AMS1}. So the present compound YbFe$_{2}$Al$%
_{10}$ offers a fertile ground to study the low temperature correlation
among 3$\mathit{d}$ Fe along with high temperature fluctuations due to 4$%
\mathit{f}$ Yb moments.

\subsection{II. \textbf{Electrical} \textbf{resistivity}}

The temperature dependence of resistivity $\rho $($T$) in YbFe$_{2}$Al$_{10}$
was obtained in the temperature range 1.8$\leq T\leq 50$ K in three applied
magnetic fields for the first sample and in an extended temperature down to
0.4 K in zero field and 9T for the second sample. \ The magnetic
contributions to the resistivity were obtained by subtracting the
resistivity of LaRu$_{2}$Al$_{10}$ as a non magnetic reference. The
resulting normalized [ $\rho $($T$)/$\rho (40K)]$ data are shown in Fig.5.
The $\rho $(\textit{T})/$\rho (40K)$ data exhibits a logarithmic divergence
in the intermediate temperature regime and passes through a maximum at about
4.5 K and decreases at low temperature.The crossover from incoherent Kondo
scattering to coherent Kondo scattering behavior of the resistivity below
4.5 K can be interpreted in the framework of the Kondo impurity model with $%
\mathit{S}$\textit{\ }= 1/2\cite{DR,TAC,TAC2}.


\FRAME{ftbpFU}{3.2543in}{2.0773in}{0pt}{\Qcb{(Color online) (a) Temperature
dependence of $\protect\rho $(\textit{T})/$\protect\rho (40K)$ in various
applied magnetic fields for sample 1. (b) Temperature dependence of $\protect%
\rho $(\textit{T})/$\protect\rho (40K)$ in various applied magnetic fields
for sample 2. }}{}{fig5.eps}{\special{language "Scientific Word";type
"GRAPHIC";maintain-aspect-ratio TRUE;display "ICON";valid_file "F";width
3.2543in;height 2.0773in;depth 0pt;original-width 11.2123in;original-height
7.9096in;cropleft "0.0095";croptop "0.9093";cropright "0.9667";cropbottom
"0.0468";filename '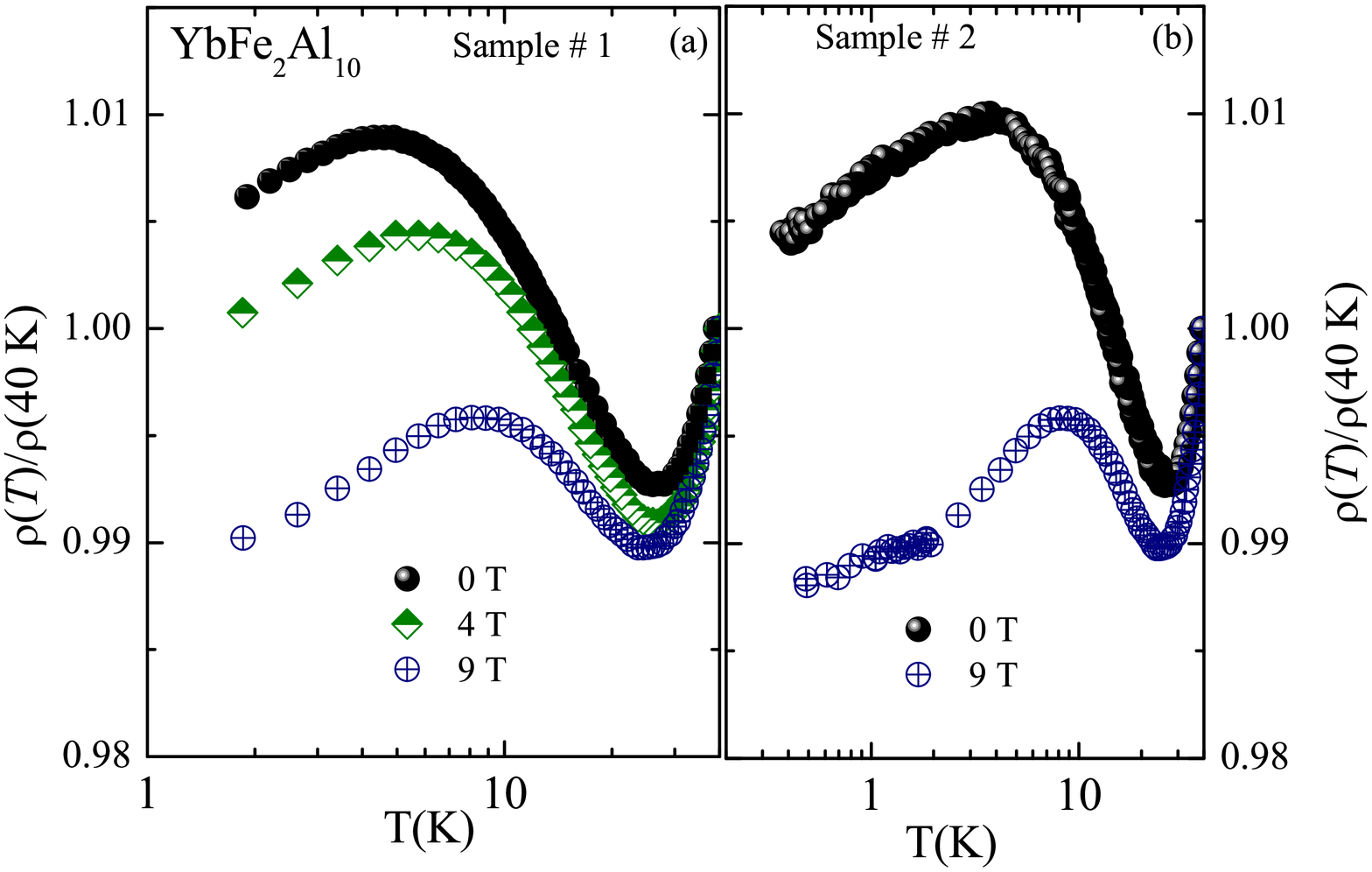';file-properties "XNPEU";}}


The field dependence of electrical resistivity is consistent with the
interpretation of Costi for dilute magnetic alloys\cite{TAC,TAC2,WF} and
could be explained as well in good agreement with our experimental data. The
observed behavior in resistivity is reminiscent of a Kondo-lattice behavior
and therefore clearly evidences the presence of strong electron correlations
among Fe ($\mathit{S}$\textit{\ }= 1/2) moments at low temperatures. These
results are consistent with magnetic susceptibility data wherein the low
temperature admixed FM correlations are dominated by Fe moments and Yb
displays an intermediate valence state.

\subsection{III. \textbf{Specific heat}}

As a further measure of the low temperature correlations, specific heat
studies on polycrystalline samples have been performed in various applied
fields in the temperature range 0.35 $\leq T\leq $ 10 K using the $^{3}$He
option of QD PPMS. It may be noted that in high magnetic fields $C_{p}$($T$)/%
$T$ slightly increases at very low temperature, which is attributed to the
high temperature part of the nuclear Schottky contributions. This part could
be described as $C_{N}$=$\beta $/$T^{2}$ (where $\beta $ depends on magnetic
field), which appears because of the Zeeman interaction and the electric
field gradient at the nuclear site responsible for lifting the degeneracy of
the nuclear energy levels. The nuclear Schottky contributions to the
specific heats in magnetic fields are subtracted\cite{ESR1} and the
resulting plot is shown in Fig. 2a in the manuscript.

\subsection{IV. $^{27}$\textbf{Al (\textit{I }= 5/2) Nuclear Magnetic
Resonance (NMR)}}

$^{27}$Al-NMR ($I=5/2$) spectra and spin-lattice relaxation measurements
have been performed using a standard \emph{Tecmag} NMR spectrometer in the
temperature range $1.8\leq T\leq 300~$K and in the field range $0.98\leq \mu
_{0}H\leq 7.27~$T. YbFe$_{2}$Al$_{10}$ hosts five inequivalent Al sites but
different Al sites are rather equal in their magnetic environment, which are
very similar to our previous NMR\ results on the structural homologue YFe$%
_{2}$Al$_{10}$\cite{PK}. There are no sharp features assigned to the first
order quadrupolar transitions and no appreciable shift observed in the $%
^{27} $Al-NMR line (Fig.\ 3 of the manuscript), but instead a broadening of
the central transition with decreasing temperatures is found. The line width
(FWHM) increases with decreasing temperature and scales with the bulk
susceptibility yielding a Curie-Weiss like behavior. At the lowest $T$ and
in small magnetic fields the scaling of FWHM with $\chi (T)$ breaks down,
which is in-line with the expected behavior at the onset of electronic
correlations. Furthermore, the narrow central transition evidences high
purity of the sample studied here and indicates the absence of onsite
disorder and Al-Fe site exchange.

The recoveries of longitudinal magnetization at time\textit{\ }$\mathit{t}$, 
$\mathit{M}_{z}$($\mathit{t}$) after the saturation pulse in the temperature
and field range of the present investigation were fitted with a single
component appropriate for $\mathit{I}$= 5/2 nuclei 1-$\mathit{M}_{\mathit{z}%
} $($\mathit{t}$)/$\mathit{M}$($\infty $) =0.0291e$^{-t/T_{1}}$+0.178e$%
^{-6t/T_{1}}$+0.794e$^{-15t/T_{1}},$where \textit{M}($\infty $) is the
equilibrium magnetization\cite{PK2, AN}.

\subsection{V. \textbf{Hard x-ray photoemission spectroscopy(HAXPES)}}

In order to gain further insights into the electronic states and to
determine the valence state of Yb in YbFe$_{2}$Al$_{10}$, hard x-ray
photoemission spectroscopy (HAXPES) with $h\nu =6.5$ keV were performed at
BL12XU of SPring-8, Japan. The HAXPES spectra were taken by using a
hemispherical analyzer (MB Scientific A1-HE) and the overall energy
resolution was set to about 0.2 eV. Clean surface of the sample was obtained
by fracturing \textit{in situ} and the spectra were measured at 50 K. The
binding energy of the spectra was calibrated by the Fermi edge of a gold
film.


\FRAME{ftbpFU}{3.154in}{2.3557in}{0pt}{\Qcb{Wide scan of YbFe$_{2}$Al$_{10}$
measured at 50 K. No O 1$s$ and C 1$s$ were observed. }}{}{fig6.eps}{\special%
{language "Scientific Word";type "GRAPHIC";maintain-aspect-ratio
TRUE;display "ICON";valid_file "F";width 3.154in;height 2.3557in;depth
0pt;original-width 11.3757in;original-height 7.9468in;cropleft
"0.0820";croptop "0.9192";cropright "0.8968";cropbottom "0.0501";filename
'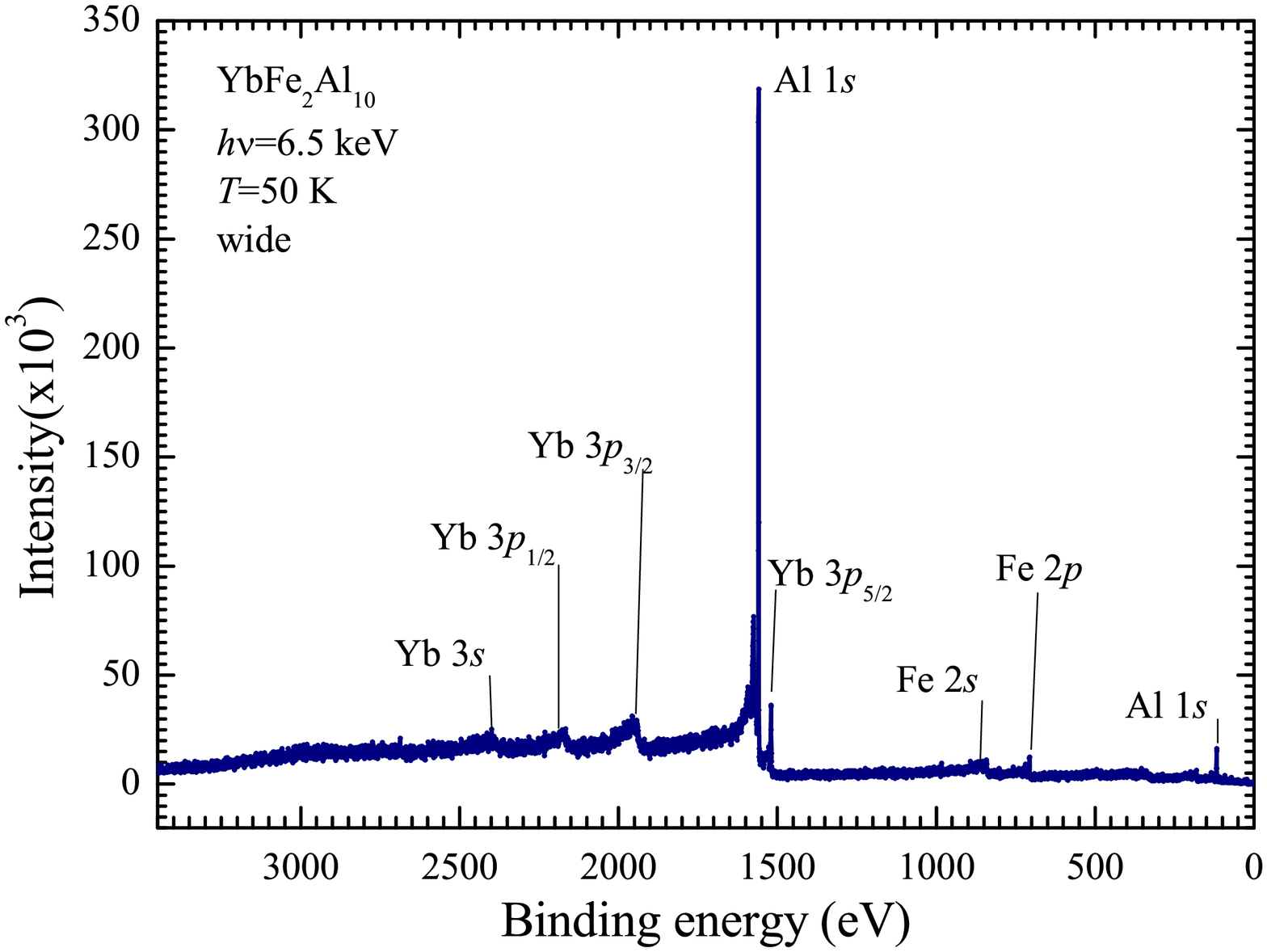';file-properties "XNPEU";}}


A full-range energy HAXPES spectrum of YbFe$_{2}$Al$_{10}$ revealing the
well-resolved three elemental contributions from the title compound is shown
in Fig. 6. We note the absence of any extraneous contributions such as
oxygen or elemental carbon.


\FRAME{ftbpFU}{3.1107in}{2.7406in}{0pt}{\Qcb{Yb 3$d$ spectrum of YbFe$_{2}$Al%
$_{10}$ measured at 50 K. The Al $1s$ peak with its plasmon peaks are also
indicated.}}{}{fig7.eps}{\special{language "Scientific Word";type
"GRAPHIC";maintain-aspect-ratio TRUE;display "ICON";valid_file "F";width
3.1107in;height 2.7406in;depth 0pt;original-width 10.1676in;original-height
7.9096in;cropleft "0.0964";croptop "0.9061";cropright "0.8980";cropbottom
"0";filename '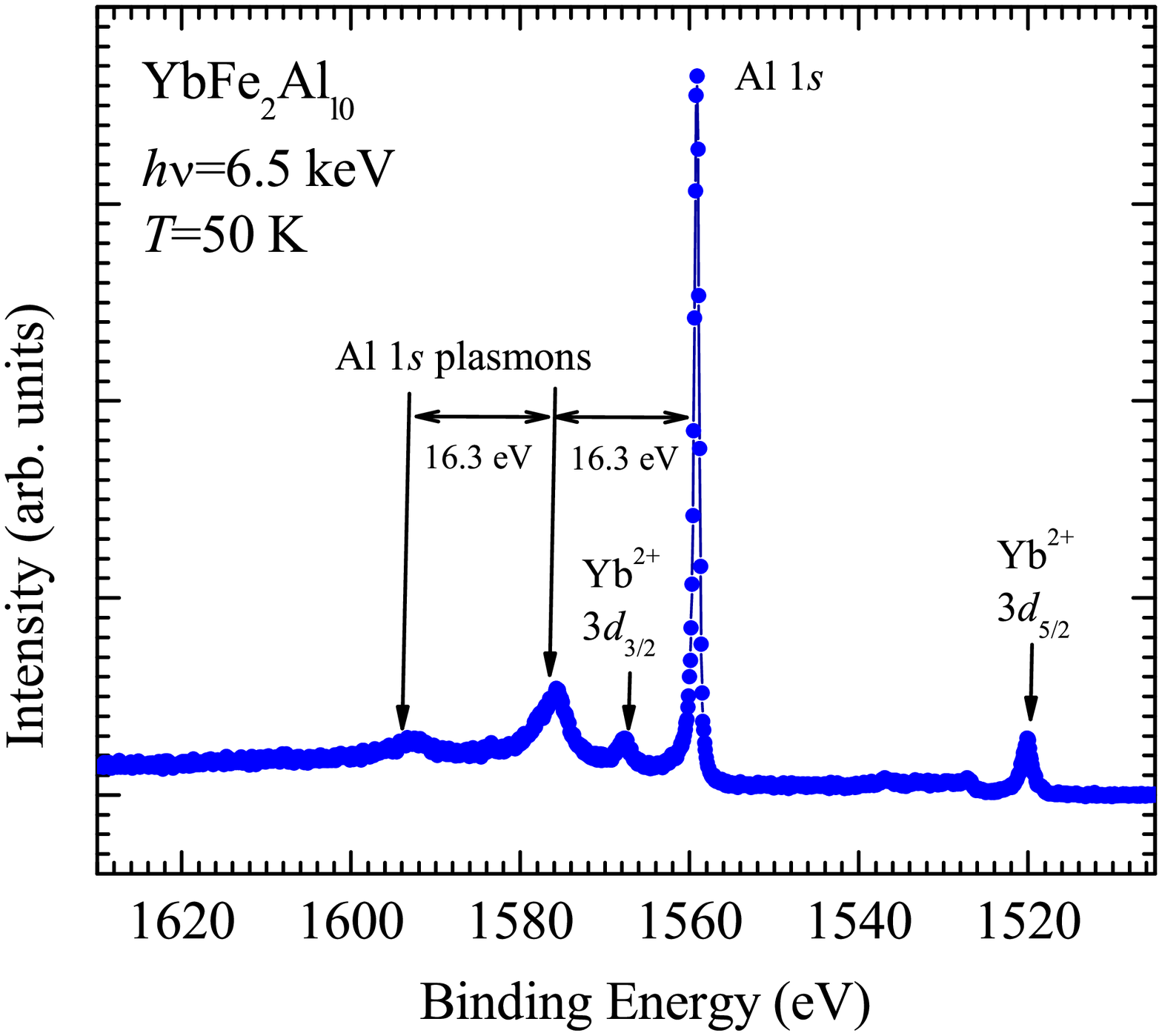';file-properties "XNPEU";}}


Figure 7 depicts the Yb $3d$ spectra of YbFe$_{2}$Al$_{10}$ measured at 50
K. The Yb 3$d$ spectrum is split into 3$d_{5/2}$ region at 1515-1540 eV and 3%
$d_{3/2}$ region at 1565-1580 eV due to the spin-orbit interaction. A strong
Al 1$s$ peak can be observed around 1560 eV and its plasmon peaks are
visible at 16.3 eV higher binding energies and multiples thereof.


\FRAME{ftbpFU}{3.6409in}{2.9006in}{0pt}{\Qcb{The Yb 3$d_{5/2}$ spectra of
YbFe$_{2}$Al$_{10}$ measured at 50 K and its simulation consisting of the Yb$%
^{2+}$ and Yb$^{3+}$ $3d_{5/2}$ multiplets and their plasmon satellites as
well as an integral backgound.}}{}{fig8.eps}{\special{language "Scientific
Word";type "GRAPHIC";maintain-aspect-ratio TRUE;display "ICON";valid_file
"F";width 3.6409in;height 2.9006in;depth 0pt;original-width
10.0024in;original-height 7.9096in;cropleft "0.0317";croptop
"0.9597";cropright "0.9865";cropbottom "0";filename
'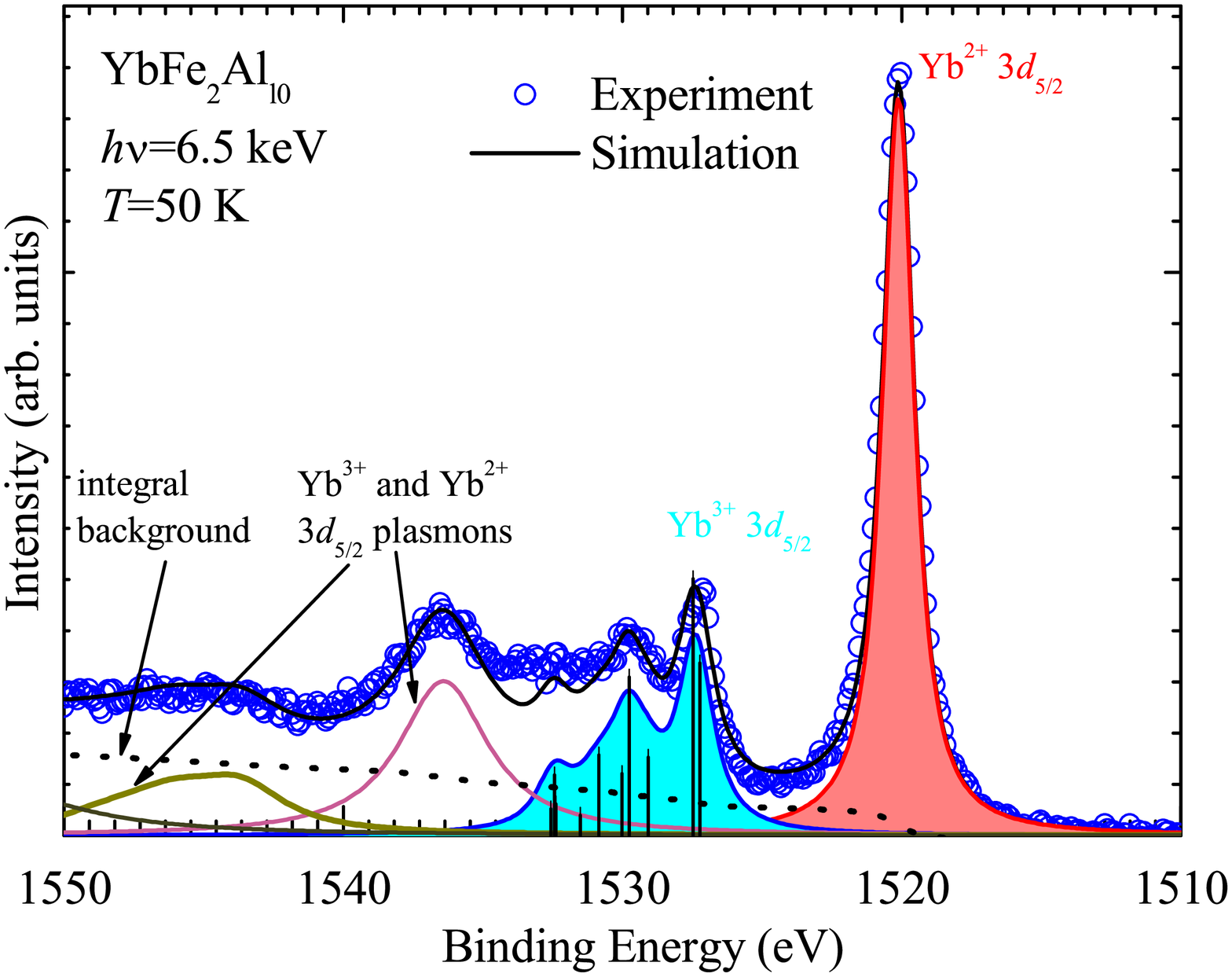';file-properties "XNPEU";}}


Figure 8 shows in more detail the Yb $3d_{5/2}$ spectra of YbFe$_{2}$Al$%
_{10} $ measured at 50 K. Here, we will evaluate only the 3$d_{5/2}$ part of
the Yb 3$d$ spectra because the tail of the enormous Al 1$s$ peak and its
plasmon structure are overlapping with the 3$d_{3/2}$ region. The Yb$^{2+}$
component is observed as a prominent peak at 1520 eV and the Yb$^{3+}$
component shows up at 1525-1535 eV as a multiplet structure arising from the
Coulomb interaction between the 3$d$ and 4$f$ holes in the electronic
configuration of the 3$d^{9}4f^{13}$ final states. The structures at higher
binding energies can be attributed to the plasmon satellites of these Yb
core levels. Coexistence of the Yb$^{2+}$ and Yb$^{3+}$ structures indicates
directly the intermediate valence states of Yb in YbFe$_{2}$Al$_{10}$.

To extract a number for the Yb valence in YbFe$_{2}$Al$_{10}$ we performed a
simulation of the spectrum by taking into account not only the Yb$^{3+}$
line and the Yb$^{2+}$ multiplet structure, but also their respective
plasmon satellites, the relative intensities and energy positions of which
were calibrated using the Al 1$s$ and its plasmons. Including also the
standard integral background, we were able to obtain a satisfactory
simulation of the experimental spectrum, see Fig.8. The Yb valence is then
estimated to be 2.38. We note that a similar HAXPES study was performed
recently on the quantum critical intermediate valent compound $\beta $-YbAlB$%
_{4}$ by M. Okawa \textit{et al}\cite{MOX}. There, the much higher
prevalence of the magnetic Yb$^{3+}$ state is consistent with the heavy
fermion ground state in $\beta $-YbAlB$_{4}$, whereas in YbFe$_{2}$Al$_{10}$
the Yb$^{3+}$ is found to play a much more subdued role.

\bigskip

*pkhuntia@gmail.com

\end{document}